\documentclass[manuscript]{aastex631}

\shorttitle{TWO-SIDED LOOP JET IN A FILAMENT CHANNEL}
\shortauthors{Yang et al.}

\graphicspath{{./}{figures/}}

\begin{document}

\title{Two-sided Loop Solar Jet Driven by the Eruption of a Small Filament in a Big Filament Channel}

\author[0000-0003-3462-4340]{Jiayan Yang}
\affiliation{Yunnan Observatory, Chinese Academy of Sciences, P.O. Box 110, Kunming 650011, China}

\author{Hechao Chen}
\affiliation{School of Physics and Astronomy, Yunnan University, Kunming 650500, China}

\author{Junchao Hong}
\affiliation{Yunnan Observatory, Chinese Academy of Sciences, P.O. Box 110, Kunming 650011, China}

\author{Bo Yang}
\affiliation{Yunnan Observatory, Chinese Academy of Sciences, P.O. Box 110, Kunming 650011, China}
\affiliation{Yunnan Key Laboratory of Solar Physics and Space Science, Kunming 650011, China}

\author{Yi Bi}
\affiliation{Yunnan Observatory, Chinese Academy of Sciences, P.O. Box 110, Kunming 650011, China}
\affiliation{Yunnan Key Laboratory of Solar Physics and Space Science, Kunming 650011, China}

\begin{abstract}

Similar to the cases of anemone jets, two-sided loop solar jets could also be produced by either flux emergence from
the solar interior or small scale filament eruptions.
Using the high-quality data from the {\it Solar Dynamic Observatory}
({\it SDO}), we analyzed a two-sided loop solar jet triggered by the eruption of a small filament in this paper. The jet was
occurred in a pre-existing big filament channel. The detailed processes involved in the small filament eruption,
the interaction between the erupted filament and the big filament channel, and the launch of the two-sided loop jet are presented.
The observations further revealed notable asymmetry between the two branches of the jet spire,  with the northeastern
branch is narrow and short, while the southern branch is wide and long and accompanied by discernible untwisting motions.
We explored the unique appearance of the jet by employing the local potential field extrapolation to calculate the coronal
magnetic field configuration around the jet. The photospheric magnetic flux below the small filament underwent cancellation
for approximately 7 hours before the filament eruption, and the negative flux near the southern foot-point of the filament
decreased by about 56 percent during this interval. Therefore, we proposed that the primary photospheric driver
of the filament eruption and the associated two-sided loop jet in this event is flux cancellation rather than flux emergence.

\end{abstract}

\keywords{Solar activity (1475)  --- Solar filament eruptions (1981) ---
Solar magnetic fields (1503)}

\section{Introduction} \label{sec:intro}
Solar jets are ubiquitous transient phenomena occurred in the solar atmosphere. They are collimated plasma beams
expelled along open field lines or far-reaching coronal loops, with a width range of $10^2 \sim 10^5$ kilometres \citep{
shimojo00, paraschiv15}. With the aids of advanced ground-based and space-borne solar telescopes, solar jets now
could be observed across various wavelengths, from H$\alpha$ \citep[where they were called as surges;][]
{roy73,jibben04, liu08, yang12a, yang12b, li17}, extreme-ultraviolet (EUV) wavelengths \citep{nistico09,chen12, moschou13,
zhang16}, to X-rays \citep{shibata92, shimojo96, paraschiv15}. Actually, solar jets are often observed at different wavebands
simultaneously \citep{jiang07, chen08, tang21, yang23b}.

Morphologically, solar coronal jets are categorized into two types \citep{shibata94a, shen21}: straight anemone jets
and two-sided loop jets. Anemone jets have been discussed extensively in past decades. \citet{sterling15} refer to them as
``single-spire jets'' since they only exhibit a single spire. The spire of anemone jet often vertical to the solar surface approximately,  
and a bright point often occurred at the edge of the jet base. 
In contrast, two-sided loop coronal jets are bidirectional jets. They consist of two roughly antiparallel
spires, which usually develop symmetrically from the eruption source region and extend horizontally to the solar surface.
The trigger mechanism of both anemone jets and two-sided loop jets is traditionally explained by the emerging-flux
model \citep{shibata94a, yokoyama95}. In this model, a magnetic bi-pole emerging from the solar interior undergoes
magnetic reconnection with the ambient pre-existing open field or far-reaching loops, resulting in the formation of jet spires
along that open (or far-reaching) field. If the ambient coronal field is oblique or vertical to the solar surface, the produced jet
will be an anemone jet; while if the overlying coronal filed is horizontal, a two-sided loop jet will be initiated.

Early magnetohydrodynamic (MHD) simulation of emerging-flux model is 2D \citep{yokoyama95}. Subsequently,
 \cite{moreno08} presented the first 3D numerical experiment of the model, and \cite{moreno13} extended the work further.
 Observations also revealed that some anemone jets were triggered by the emergence of magnetic flux and thus validated
the emerging-flux model \citep{shibata94b, zhang00, chen08, li15}. Furthermore, \cite{moreno13} and some other MHD
simulations illuminated that a parasitic bipolar magnetic field emerging from the solar interior would naturally evolve into a
sheared arcade. This is consistent with observations that flux emergence is often accompanied by photospheric shear motions,
allowing for the transfer of magnetic helicity to the chromosphere and corona \citep{xu22}. This sheared arcade would subsequently
reconnect with the pre-existed ambient field, and produce a so-called
blowout jet \citep{moore10}. Therefore, these numerical experiments provided an improved version of the original emerging-flux
model. Accordingly, new observations also reported to support the model. Recently, \cite{schmieder22} reviewed the research
results of surges and jets observed by the Interface Region Imaging Spectrograph \citep[IRIS;][]{depontieu14}. Several cases
discussed in their review reported the jets produced by flux emergence, aligning well with the improved emerging-flux model.
One of these cases was presented by \cite{ruan19}. Utilizing IRIS spectrographic observations, they reported bi-directional flows
with velocities of nearly $\pm200$ km s$^{-1}$ in an active region, and attributed these flows to the outflows of magnetic
reconnection occurring between emerging flux and a long twisted loop. The magnetic reconnection produced a twisted jet as well.
\cite{joshi20b} analyzed another twisted jet with the help of IRIS spectrographic observations. They have displayed the detailed
process of how the twist transferred from a flux rope located in the reconnection site to the jet. Particularly, \cite{joshi20a}
studied six recurrent EUV jets observed by SDO/AIA and IRIS simultaneously. They found each of these jets presented a
double-chambered structure with cool and hot emissions in each vault, corresponding to the cool and hot loop regions predicted
by the emerging-flux models of \cite{moreno08} and \cite{moreno13}.

On the other hand, numerous observations have revealed that the generation of anemone jets does not always involve the
emergence of  magnetic flux, but is closely associated with the eruption of filaments/mini-filaments in the source regions
\citep{hong11, yang12a, sterling15, hong16, shen17, duan22}. Specifically, jets triggered by mini-filament eruptions typically
manifest as blowout jets \citep{moore10}, and magnetic flux cancellation rather than flux emergence often occurs before
the eruption of mini-filaments and the launch of jets \citep{adams14, panesar16, joshi18, panesar18, duan19, chen20, yang20}.
Based on the observations of 20 randomly selected solar coronal jets in polar coronal holes, \cite{sterling15} found that each
of them was driven by the eruption of a mini-filament, and proposed a revised picture to interpret the production of jets.
Simulations also support the scenario that filament/mini-filament eruption is the source of coronal jet \citep{wyper17,wyper18}.
Using ultrahigh-resolution 3D magnetohydrodynamic simulations, \citet{wyper17} demonstrated the production process of
a coronal jet through the eruption of a mini-filament or flux-rope structure embedded in open unipolar field, and its subsequent
evolutionary stages. \citet{wyper18} further studied in detail three realizations of the model to explain different observational
features.

As we mentioned above, theory models for producing solar jets are applicable to both anemone jets and two-sided loop
jets. However, comparing with anemone jets, the observational studies of two-sided loop jets are insufficient till now.
Some cases have shown magnetic flux emerging from below the photosphere, reconnecting with a pre-existing overlying
horizontal field, and producing a two-sided loop jet \citep{kundu98, jiang13, zheng18, tan22}, thus supporting the
emerging-flux model. Beside these, there are also some other high-resolution observations revealed that the eruption
of mini-filaments plays an important role in the formation of two-sided loop jets, aligning with the perspective of
\cite{sterling15} for anemone jets. \citet{tian17} presented observational analysis of two successive two-sided loop jets
which were produced not by the emergence of magnetic flux but by the reconnection of two filamentary threads.
 \citet{sterling19} presented an event to show strong evidence that a two-sided loop jet resulted from an erupting
mini-filament, and the magnetic trigger of the eruption is apparently flux cancellation. The examples from \citet{shen19}
and \citet{yang19} further advanced our understanding of two-sided loop jets. Beside confirming the role of mini-filaments
in triggering two-sided loop jets, \citet{shen19} revealed that the birth of the two-sided loop jet involved two reconnection
processes, while \citet{yang19} inferred that the mini-filament could reform at the same neutral line after its first eruption
and produce recurrent two-sided loop jets. Particularly, \cite{yan21} reported that  continuous magnetic cancellations could
produce a series of unidirectional and bidirectional jets with the eruption of a hot channel rather than a mini-filament.

The current study focuses on a two-sided loop solar jet that appeared in a big filament channel on the eastern part
of the solar disk on March 22, 2012.  This event provides a good opportunity to investigate the triggering mechanism
of two-sided loop solar jets from a perspective of observation,  which has been relatively scarce until now. Moreover, this
jet deviates from the typical symmetrical spires seen in many two-sided loop jets, exhibiting obvious asymmetric jet spires.
Through  an examination of the evolution  of photospheric magnetic field and the eruption process of the jet, we aimed to
identify the trigger of the jet and the reason behind its distinctive appearance. The paper is arranged as follows: the instruments
and the data we used are described in Section 2, Section 3 presents the main observational results along with our explanations
for the event,  and in Section 4 the summary of the event and a brief discussion are given.

\section{Instruments}\label{sec:instru}
The two-sided loop jet we reported in this paper was observed perfectly by the Atmospheric Imaging Assembly
\citep[AIA;][]{lemen12} on board the {\it Solar Dynamics Observatory} \citep[{\it SDO};][]{pesnell12}. AIA takes full-disk images
in 10 UV/EUV passbands with a pixel size of $0^{''}.6$ and a cadence of 12 seconds. We mostly utilize the Level 1.5 images
of its 7 EUV passbands here, which are 304 \AA\ (He {\sc ii}, log{\it T} = 4.7), 171 \AA\ (Fe {\sc ix}, log{\it T} = 5.8), 193 \AA\
(Fe {\sc xii}, log{\it T} = 6.2), 211 \AA\ (Fe {\sc xiv}, log{\it T} = 6.3), 131 \AA\ (Fe {\sc xxi}, log{\it T} = 7.0), 94 \AA\ (Fe {\sc xviii},
log{\it T} = 6.8), and 335 \AA\ (Fe {\sc xvi}, log{\it T} = 6.4). The 304 \AA\ and 171 \AA\ images are especially important when
analyzing the eruptions of the jet in this event, but the data from other AIA EUV passbands are also examined.
Full-disk line-of-sight (LOS) magnetograms obtained by the Helioseismic and Magnetic Imager \citep[HMI;][]{scherrer12}
on board the {\it SDO} were used to investigate magnetic field in the source region of the jet. These magnetograms
were obtained in the Fe {\sc i} absorption line at 6173 \AA\, with a spatial sampling of $0^{''}.5$ pixel$^{-1}$,
a cadence of 45 seconds, and a noise level of 10 G. All the data are processed by using standard
software programs in the SolarSoftWare ({\it SSW}), and differentially rotated to a reference time close to the event.
In addition, to exhibit the site and appearance of the filament channel, we also use H$\alpha$ line-center images obtained
by the Global Oscillation Network Group \citep[GONG;][]{harvey96} at the National Solar Observatory (NSO). The pixel size of
the GONG H$\alpha$ images is $1^{''}$, and the time resolution is 1 minute.

\section{Observational Results}\label{sec:result}

On March 22, 2012, a two-sided loop solar coronal jet was observed by SDO/AIA in the eastern part of the solar disk,
near the equator. The jet wasn't associated with any recorded flares or coronal mass ejections (CMEs).
Figure 1 displays the location and the general appearance of the jet and its source region. It's worth noting that
panel (d) of Figure 1 was captured on February 23, 2012, while other images in this figure were obtained on March 22.
As showed by the AIA 171 and 304 \AA\ images (panels (a) and (b)), the jet located at the northeastern end
of a large curved dark structure. Upon careful examination of GONG H$\alpha$ observations, we found that a big
quiescent filament (QF) was presented in the same region approximately one solar rotation cycle earlier (panel (d)),
and the shape of QF resembles the dark structure seen in panels (a) and (b). However, on the date when the two-sided
loop jet occurred, the quiescent filament QF had already disappeared, leaving behind only some fragmented
remnants, which are pointed out by the white arrows in the GONG H$\alpha$ image (panel (c)). Therefore, the dark structure
we see in AIA 171 and 304 \AA\ images should be a filament channel (FC). A simultaneous zoomed-in 171 \AA\ image is
inserted in panel (a) to provide a clearer view of the jet's appearance. As depicted, the jet consists of
a bright jet base and two almost anti-parallel spires. However, unlike typical two-sided loop solar coronal jets,
where the two branches of the jet spire are usually symmetrical, the jet presented here exhibits asymmetric branches
of the jet spire, i.e., the northeastern spire is short and narrow, while the southern spire is long and wide.

Further examination of the source region of the jet (marked by white boxes in panels (b) and (c)) reveals
the presence of a small filament F before the jet occurred (panels (e-g)). In the 04:40:20 UT
AIA 304 \AA\ image  (panel (f)), F presented itself as a hook-like dark structure, with the length of about
$47^{''}.54$ or $3.447 \times 10^4$ km. However, it is not easy to identify this small filament in AIA 171 \AA\
and GONG H$\alpha$ images at the same time. F became more apparent a few minutes later in these wavebands (see panels
(e) and (g)), and its shape and the length are differed from what was observed in the 04:40 UT AIA 304 \AA\ image
(the outline of F in this image is overlaid as black curves in panels (e) and (g)). By superimposing the contours of an
almost simultaneous HMI magnetogram onto the 304 \AA\ image (panel (f)), it was deduced that the northern foot-point
of filament F rooted in the positive magnetic field region, while its southern foot-point rooted in the negative
region. Additionally, the photospheric magnetic fields beneath F's southern foot-point appeared to be highly
sheared. This statement is further supported by the HMI magnetogram superposed with F's outline in panel (h).

Figure 2 presents the eruption process of the small filament F and the following two-sided loop jet.
As indicated by the black rectangle in panel (c1), the zoomed-in AIA 304 and 171 \AA\ images of the jet's source
region are presented in panels (a1-b4). The filament F is identifiable in these images. It is noticed that,
at 04:40:20 UT, a patch of brightening existed beneath the south end of F already (panel (a1)).
When tracking AIA and HMI observations before the time, it is inferred that the brightening had been present for about
one day since 04:00 UT on March 21, and the photospheric magnetic field in the region has experienced continuous
flux cancellations. Therefore, we suppose that the brightening would be a coronal bright point (CBP), which usually
appeared in quiet-Sun regions and coronal holes, associated with opposite magnetic polarities in the photosphere,
and had an average lifetime of about 28 hours in EUV images \citep{hong14}. However, the CBP did not contain any
filament until 04:00 UT on March 22. In the accompanying animation of Figure 2 which starts at 04:00 UT on March 22,
it is noticed that F was unrecognizable at the very beginning, and then it started to form a few minutes later. 
Consistent with panel (a1), at 04:40 UT F could be identified clearly at the animation. Therefore, although AIA images did not 
got sufficient resolution to reveal either the exact time or the detailed process of F's formation, we could infer that F
formed between 04:00 UT and 04:40 UT on March 22.

The new-formed small filament F was obvious and stable at 04:40:20 UT as showed in the  AIA 304 \AA\ image (panel (a1)),
but it soon started to lift soon. When superimposing the original position of F (blue curves, obtained from panel (a1)) to other
304 \AA\ images in the top row, it is clear that F erupted southeastwardly, approaching the big filament channel FC. The CBP
also became brighter and more expanded, evolved into a small flare eventually. The subsequent movement of the erupting
filament is more clearly visible in AIA 171 \AA\ images (panels (b1)-(b4)). As pointed out by the white arrows in panels
(b1) and (b2), F erupted towards the pre-existing filament channel FC, and its appearance changed a lot compared to
the original form in AIA 304 \AA\ image (blue curve). The overlying magnetic field of erupting F would reconnect
with the magnetic field of FC, resulting in the launch of a jet. Dark filament plasma was observed to be ejected into FC, and some
brightenings occurred at the interface of them, which were likely signatures of the magnetic reconnection (panels (b3) and (b4)).
Then, bright mass flow streamed from a bright point near the flare and flowed southwardly into FC.

The observed bright mass flow indicates the initiation of the jet, while the complete appearance and development of the
jet were displayed in panels (c1)-(c4) of AIA 171 \AA\ images with a larger field of view (FOV) . In these pictures,
it is evident that the bright jet plasma flowed both southwardly and northeasterly, forming two roughly antiparallel jet
spires, and the small flare associated with the filament eruption evolved into the jet bright point (JBP). Based on
the appearance of the jet and the trajectory of F's eruption, it can be deduce that the jet was a two-sided loop solar jet
triggered by the eruption of the small filament F. Therefore, this event supports the picture proposed by
\cite{sterling15}, confirms once again that two-sided loop solar jets could also be produced by the eruption of small
scale filaments. It is consistent with recent observations of two-sided loop jets \citep{shen19, sterling19, yang19}.
It is obvious that the northeastern spire of the jet is short and narrow, resembling a standard jet, while the southern
spire is long and wide, resembling a blowout jet \citep{moore10}. In this sense, the jet is particular, since most two-sided
loop jets typically have roughly symmetrical spires \citep{jiang13, sterling19, yang19}. Additionally, the northeastern
spire of the jet appears bright, indicating it is composed mainly of hot plasma, while the southern spire is a mixture of cold
and hot mass, and exhibits noticeable untwisting motion (see the accompanying animation).

Panel (d) of Figure 2 displays the normalized AIA 304 and 171 \AA\ light curves measured in the white boxes in
panels (a4), (b3), and (c4). The light curves show the evolution of the intensity of the CBP (or the flare, or JBP). It
is noticed that the brightness of the patch began to increase sharply at around 04:45 UT, indicating the
beginning of F's eruption. Both the 304 and 171 \AA\ light curves reached their first peak simultaneously at approximately
04:49 UT, as indicated by the vertical dotted line. This peak should correspond to the maximum intensity of the small flare
produced by the filament eruption.  Subsequently, the intensities of the 304 and 171 \AA\ images in the region decreased
for a while, then increased again and reached the their maximum value. This second stage of increasing of intensity may
indicate the launch of the jet, and the maximum of the light curve corresponds to the maxima of the JBP caused by the jet.
Note that the time of maximum intensity differs when measured in different wavebands: the 171 \AA\ light curve reached
its peak at about 05:11 UT, as indicated by the vertical dashed line, while the 304 \AA\ light curve  reached its
peak at about 05:07 UT, a few minutes before the maximum of the 171 \AA\ light curve.

To analyze the evolution of the jet in detail, time-slice plots were constructed from AIA 304, 171, and 131 \AA\
images and displayed in Figure 3. In this figure, panels (a)-(c) show the time-slices constructed along
slit S1, while panel (d) presents the time-slice along slit S2.  Slit S1 is a curved line segment along the jet spire, pointing
from the northeast to the south, as shown by the white arrows in panels (c3) and (c4) of Figure 2. Slit S2,  on the
other hand, is a straight line segment roughly vertical to the southern branch of the jet spire.
In panel (b) of Figure 3, the horizontal dashed line  represents the locations of the demarcation point of the jet spire at
different times, corresponding to the blue asterisks in Figure 2. Therefore, in this image, the region above the
dashed line describes the southward ejection of the jet, while the region below the line shows the northeastern ejection
of the jet. The ejections along the two branches of the jet spire show notable differences. The southward ejections
are generally more vigorous, with a larger amount of plasma being ejected. Furthermore, the southward ejections are
steady and continuous, consisting of both dark and bright components, indicating the ejection of
cold filament mass into the southern branch of the jet spire. All three time-slices along S1 (panels (a-c)) show that the mass
ejection along the southern branch of the jet spire can be divided into two stages: a slow ejection stage with an average projected
speed of about $18.09$ km s$^{-1}$, as measured in 171 \AA\ time-slice, and a fast ejection stage with an average projected
speed of about $165.05$ km s$^{-1}$. This behavior is similar to the eruption of a filament, which aligns with what we
inferred from Figure 2 that the jet was triggered by a small filament eruption. \citet{zhang21} also found that the kinetic evolution
of an anemone jet triggered by a mini-filament eruption can be divided into a slow rise phase and a fast rise phase, and they
suggested that these two phases may correspond to the magnetic reconnections at the breakout current sheet and the flare
current sheet, respectively. According to the 304 \AA\ time-slice (panel (a)), the fast ejection of the jet started at about
05:00:29 UT, as indicated by the vertical dashed lines in the figure. On the other hand, the northeastern ejections are
intermittent and primarily consist of the bright component. Three episodes of intermittent northeastern mass ejection
are marked by red dotted lines in panel (b), with measured average speeds of $179.32$ km s$^{-1}$, $97.53$ km s$^{-1}$,
and $171.52$ km s$^{-1}$, respectively. Indeed, the speeds of the intermittent northeastern ejections are roughly comparable
to the fast ejection of the southern branch. However, the two-stages mass ejection was indistinguishable in the northeastern
branch of the jet spire,  possibly because the ejection was weak there. From the time-slice plots, the  presence of the
cold filament mass in the northeastern ejections may not be apparent, but the accompanying AIA 171 \AA\ animation
shown in Figure 2 reveals that, although the majority of F's mass was ejected into the southern branch of the jet spire, a small
portion of the dark filament mass was also ejected into the northeastern spire of the jet.

AIA animation accompanying Figure 2 showed that the southern branch of the jet spire underwent a significant untwisting
motion. This character was even more apparent in the time-slice plot along slit S2 (panel (d)). From the time-slice,
it was evident that some complicated transverse motions took place in the jet's southern spire soon after the fast
ejection of the jet. However, the width of the jet spire did not show a significant change, suggesting that the jet spire
experienced only rotational motion without expansion. Untwisting motion is a common character in blowout jets
\citep{hong13, moore15, li19, chen21}, and it has been explained as the release of the twist initially stored in the twisted
filament's magnetic field through magnetic reconnection with the ambient coronal magnetic field. Two episodes of the untwisting
motion were traced by the dotted white lines in panel (d), with linear fitting velocities of $295.37$ km s$^{-1}$ and
$211.46$ km s$^{-1}$. These rotational velocities were relatively large compared to many previous studies
\citep{zhang14, moore15, yang23b}, but still within a reasonable range. The time-slice further showed that the rotational
direction  of the southern spire of the jet was along the forward direction of S2, from the east to the west, according to
the white arrow in panel (c4) of Figure 2. This direction is consistent with what we see in the AIA 171 \AA\ animation.
Therefore, the southern spire of the jet experienced an anticlockwise rotation when viewed from JBP.

Figure 4 and its accompanying animation displayed the situation of the photospheric magnetic field before and during the
eruptions of the filament F and the following two-sided loop jet. It is revealed that, in the source region of the jet (as
indicated by the black circles in the top half part of Figure 4), the positive and negative photospheric magnetic polarities
moved toward and canceled to each other since 21:30 UT on March 21. After 01:00 UT on March 22, the cancellation occurred
mainly in the region indicated by the white ovals in the figure and the animation. When superposing the
outline of the small filament F obtained from the 04:40 UT AIA 304 \AA\ image onto the simultaneous magnetogram showed
in panel (a8), it is clear that the region in the white oval is where F's southern foot-point rooted. HMI LOS magnetograms revealed
that the negative polarity patch of the photospheric magnetic field (the black patch) continuously moved to northwestward,
while the positive polarity patch (the white patch) moved southeastward. After several hours of movement and cancelation,
the photospheric magnetic field below F's southern foot-point became highly sheared (panel (a8)), causing the southern
part of the filament to raise first, followed by the entire filament's eruption.

To demonstrate the magnetic flux cancellation in the source region of the jet more intuitively, we present the change
of the unsigned positive (red curves) and negative (black curves) magnetic fluxes in panels (b) and (c) respectively. In
panel (b), the flux curves are calculated within the region indicated by the black circles, while in panel (c), the flux
curves are calculated within the region indicated by the white ovals. It is showed that, the negative magnetic fluxes in
both regions exhibit a nearly monotonic decrease, with the exception of a short period from 00:00 UT to 01:00 UT when a
small patch of negative polarity moved into the black circle region from the south, as shown in the accompanying animation.
From 21:45 UT of the previous day to 04:45 UT of March 22, the negative magnetic fluxes around F's southern foot-point
decreased for about 7 hours before its eruption (indicated by the blue vertical lines), suggesting that the magnetic flux
cancellation occurred for several hours prior to the eruption. On the other hand, the positive magnetic flux only showed a
slight decrement from 23:00 UT on March 21 to 01:00 UT on March 22, and changed little during other measured periods.
Checking the HMI animation carefully, we realize that isolating the positive magnetic field into these regions is more
difficult compare to the negative field because its dispersed nature. Convergence motion frequently brought some positive
magnetic polarities moved into the regions, thereby enhancing the calculated positive fluxes. Additionally, it appears that
some positive polarities emerged in the regions during the flux measurements, further enhancing the positive fluxes.
However, even with the inclusion of these extra fluxes, the positive fluxes shown in panels (b) and (c) do not increase
significantly, implying cancellation with the negative fluxes at their interface.Therefore, we suspect that the eruptions
of the small filament F and the subsequent two-sided loop jet in this event are driven by photospheric magnetic flux
cancellation, similar to cases of many anemone jets \citep{hong11, panesar16, panesar18} and two-sided loop jets
\citep{sterling19, yang19} driven by mini-filament eruptions. The negative flux in the black circle
region decreased from $2.7 \times 10^{19}$ Mx to $1.2 \times 10^{19}$ Mx within 7 hours, thus the flux reduction is about
56\%, and the cancelation rate is about $2.1 \times 10^{18}$ Mx hr$^{-1}$. The values of the negative fluxes in the white oval
regions are smaller due to the smaller region size, but the flux reduction and cancelation rate are approximately the same as
those in the black circle region.

According to previous observations, anemone jets are usually observed to be vertical to the solar surface, while
two-sided loop jets are typically parallel to the solar surface, with two symmetrical branches of jet spire.
The two-sided loop jet we reported here was nearly parallel to the solar surface since it occurred in a filament channel,
but exhibited a distinctive appearance of asymmetrical branches of the jet spire. To interpret the asymmetry of the
two branches of the jet spire, we investigated the topology of the coronal magnetic field around the jet. Since we
lacked vector magnetic field data as the jet took place in the quiescent region of the solar disk, we employed the local
potential field extrapolation method \citep{alissandrakis81, gary89} to calculate the coronal magnetic field configuration
prior to the filament eruption, which is presented in Figure 5. In panel (a) which is the front view of
the solar disk, we adopted an AIA 171 \AA\ image near the maximum of the jet spire as the background image, and
superimposed extrapolated coronal field lines before the filament eruption at four representative positions above the jet's
spire. The coronal magnetic field lines above the northeastern branch of the jet's spire are denoted by the cyan and yellow
curves, while the coronal magnetic field lines above the southern branch of the jet's spire are denoted by the green and white
curves. In panel (b) which is the same image from a sideways perspective, the colors of the field lines match those
in panel (a). Combining these two panels, it is evident that the magnetic arches above the southern branch of the jet's
spire have large heights and spans, even far away from JBP (see the white curves). Therefore, this coronal magnetic configuration
provides abundant space for the southern spire to extend. In contrast, the situation is different for the
northeastern branch of the jet's spire. The heights and spans of the magnetic arches above the northeastern spire are relatively
small compared to those above the southern spire (see the yellow curves), thus the coronal magnetic field confines the jet
spire to a narrow flow. Furthermore, the coronal magnetic field becomes vertical not far ahead of the northeastern spire
(see the cyan curves), cutting off the jet spire. Therefore, the northeastern branch of the jet's spire is narrow and short.
Considering that the two-sided loop jet occurred in a big filament channel near its northeast end, the extrapolated coronal
magnetic configuration around the jet and the asymmetrical appearance of the jet's spire are reasonable.

\section{Summary and Discussion}\label{sec:summary}
Using the perfect observations from {\it SDO}/AIA and HMI, this paper presents an analysis of a two-sided loop solar
coronal jet. The jet occurred in a large filament channel located in the southeast part of the solar disk. AIA observations
revealed that a pre-existing small filament erupted in the source region just a few minutes before the jet occurred,
and captured the detailed processes of how the erupted small filament ejected into the adjacent filament channel and
triggered the two-sided loop jet. The observations confirmed that, besides the widely accepted emerging-flux
model \citep{shibata94a, yokoyama95}, two-sided loop solar jets can also be produced by the magnetic reconnection between
the overlying magnetic field of an erupting small filament and the ambient coronal magnetic field, just like in the cases of
anemone jets \citep{sterling15, hong16, li17, yang23a}. The observations further showed that two branches of the jet's
spire are heavily asymmetrical, with the southern branch being wide and long while the northeastern branch is
narrow and short. By investigating the coronal magnetic configuration around the jet, we suspect that this unique appearance
is due to the asymmetric coronal magnetic configuration near the end of the filament channel.

The trigger mechanism of solar coronal jets is an important issue for the studies of both anemone jets and two-sided loop jets.
While the event presented here maintains the picture of \cite{sterling15}, in general, either the emerging-flux model
\citep{shibata94a, yokoyama95, moreno08, moreno13} or the mini-filament-eruption-driven picture \citep{sterling15, wyper17,
wyper18} has been supported by many observation cases respectively, and the structures and kinematic characteristics of jets
produced by these two mechanisms did not show much difference from one to the other \citep{schmieder22, shen21}.
Therefore, it seems that in a pre-existing background field, both the direct emergence of magnetic flux from the interior
and the eruption of filament/mini-filament could effectively trigger reconnection between two magnetic systems and produce
a solar jet.

For the jets that conform to the emerging-flux model, flux emergence is always observed and regarded as a key driver
in photosphere \citep{shibata94b, zhang00, chen08, li15}. However, in the cases that jets are driven by erupting mini-filaments,
flux cancellation is typically predominant rather than flux emergence \citep{sterling15, panesar16, sterling19, yang19, yang23a}.
In some cases, it seems that both flux emergence and cancellation worked together \citep{jiang07, shen17, shen19}. In
our event, the positive magnetic flux in the source region of the jet did not change much, but the negative flux decreased
monotonously for about 7 hours before the eruptions of the small filament and the following two-sided loop jet. Therefore,
although could not exclude the contribution of flux emergence completely, we still propose that the main photospheric driver
of this two-sided loop jet is flux cancellation, consistent with previous studies on jets triggered by mini-filament eruptions.
A study by \cite{panesar16}  of 10 random solar jets in quiescent regions found that each one experienced flux reduction between
21\% and 57\% at the neutral line below a mini-filament. In another investigation of 13 random solar jets in coronal holes,
the authors found that the flux reduction ranged from 21\% to 73\%, with a calculated average cancellation rate of
$0.6 \times 10^{18}$ Mx hr$^{-1}$ \citep{panesar18}. The value of average cancellation rate is $1.5 \times 10^{18}$ Mx
hr$^{-1}$ for the quiescent region jets they studied previously \citep{panesar16}. In comparison, the active region jets
studied by \cite{sterling17} had an average cancellation rate of $15 \times 10^{18}$ Mx hr$^{-1}$. The two-sided loop jet in
our event occurred in a quiescent region, with magnetic flux reduction of 56\% and an average cancellation rate of
$2.1 \times 10^{18}$ Mx hr$^{-1}$. Therefore, it seems that the flux cancellation in this event was a little more violent than
the quiescent region jets studied by \cite{panesar16}, but was still within a reasonable range. This may be due to the small
filament in this event being slightly larger than the mini-filaments in their events, thus more free magnetic energy are
required to trigger the eruption of the filament.

Similar to filaments eruption of ordinary scale, the eruption of small scale filaments and the following coronal jets occasionally
result in flares and narrow CMEs \citep{yang20}. The jets can also interact with other magnetic structures during their
ejections. More recently, \cite{joshi23} reported two observational events of large-amplitude filament oscillations triggered
by solar coronal jets. In both events, a jet formed near one end of a large filament due to flux emergence. The ejected jet
hit the filament channel from the end, leading to large-amplitude oscillation in the filament, which supported the MHD
numerical experiments recently conducted by \cite{luna21}. In the present event, the jet also occurred near the end of a large
filament channel. However, during the jet's ejection, the filament channel contained little filament plasma, as indicated by
the GONG H$\alpha$ image in Figure 1. Therefore, the interaction between the jet and the filament channel was not observed
in this event.

In summary, this paper presents the observations of a two-sided loop solar jet in a large filament channel with asymmetrical
spires. The jet was triggered by the eruption of a small filament, and the main driver of the filament eruption and the jet was
the continuous  cancellation of positive and negative magnetic fluxes in the photosphere. Compared with anemone jets,
there is still limited research on two-sided loop jets, and more observations with high spatial and temporal resolution are
needed in the future to further understand their characteristics and underlying mechanisms.

\begin{acknowledgments}
We thank an anonymous referee for many constructive suggestions
and thoughtful comments that improved the quality of this paper.
The authors thank the {\it SDO} and GONG teams for granting free access to their internet databases.
This work is supported by the B-type Strategic Priority Program No. XDB41000000 and XDB0560000 funded by the
Chinese Academy of Sciences, the National Key R\&D Program of China (2019YFA0405000), the National Natural Science
Foundation of China under grants 12273108, 11933009, 12173084, 12073072, and 12273106, and by the ``Yunnan
Revitalization Talent Support Program" Innovation Team Project.
\end{acknowledgments}

\bibliography{ms}{}
\bibliographystyle{aasjournal}

\clearpage

\begin{figure}
\plotone{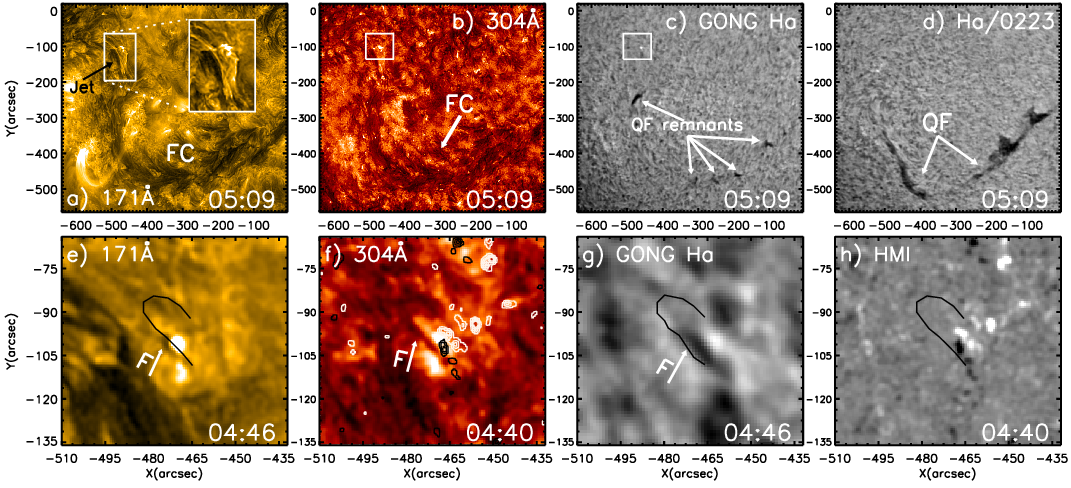}   
\caption{General appearance of the two-sided loop solar jet and its circumstances (top row) and its
source region (bottom row) showed by AIA 171 \AA, 304 \AA, GONG H$\alpha$ line-center images, and HMI LOS
magnetogram. The field of view (FOV) of the top row is $635^{''} \times 585^{''}$, while the FOV of the
bottom row is $78^{''} \times 72^{''}$, which is indicated by the white boxes in panels (b) and (c). A zoomed-in image
of the two-sided loop jet is  inserted in panel (a) to highlight it more clearly. The outline of the
axis of the small filament F obtained from panel (f) is overlaid as black curves in the other panels
of the bottom row. The LOS magnetic field measured in the 04:40:16 UT HMI magnetogram are superimposed on the AIA 304
\AA\ image (panel (f)) as black and white contours, with the intensity of $\pm 20$, $\pm 40$, $\pm 60$,
$\pm 80$, and $\pm 100$ Gauss.
\label{fig1}}
\end{figure}

\begin{figure}
\epsscale{0.95}
\plotone{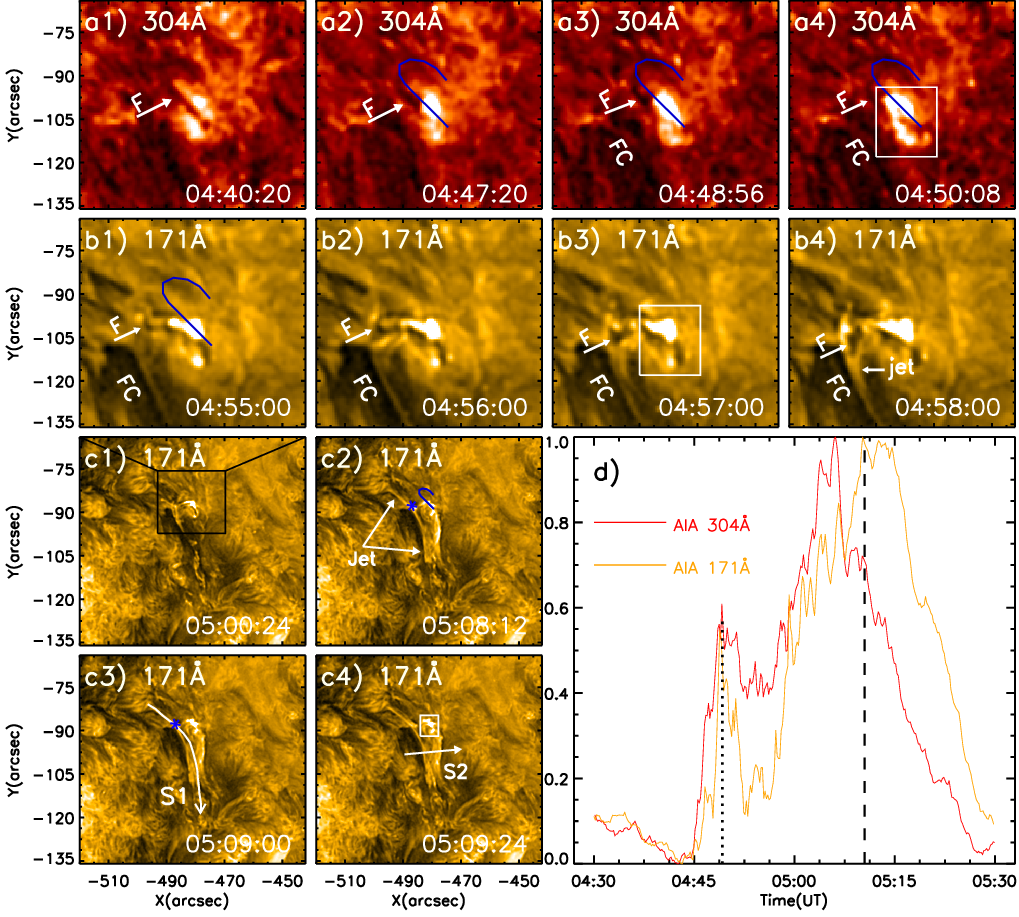}   
\caption{AIA 304 and 171 \AA\ images illustrate the eruption of the filament F (panels (a1)-(b4)) and the associated
two-sided loop jet (panels (c1)-(c4)). Some panels are overlaid with the outline of F's axis obtain from the
04:40:20 UT 304 \AA\ image as blue curves. The long curved white arrow in panel (c3) and the white arrow in panel
(c4) indicate the slit positions of the time slices shown in Figure 3, while the blue asterisks in panels (c2)
and (c3) show the approximate position of the demarcation point of the jet's spires. The black rectangle in
panel (c1) indicates the FOV of panels (a1)-(b4), which is $78^{''} \times 72^{''}$, the same as
the bottom row of Figures 1. The FOV of panels (c1)-(c4) is $260^{''} \times 240^{''}$. Panel (d)
presents the normalized light curves of AIA 304 and 171 \AA\ intensities in the region indicated by the white boxes
in panels (a4), (b3), and (c4). (An animation of AIA 171, 193, 304 \AA\ direct images and GONG H$\alpha$
line-center images is available. The animation lasts 15 seconds, has the same FOV as panels (c1)-(c4),
and covers the time from 04:00 UT to 05:29 UT. The cadence of the AIA images is 12s, while the cadence of the
GONG images is 1m.)
\label{fig2}}
\end{figure}

\begin{figure}
\epsscale{0.95}
\plotone{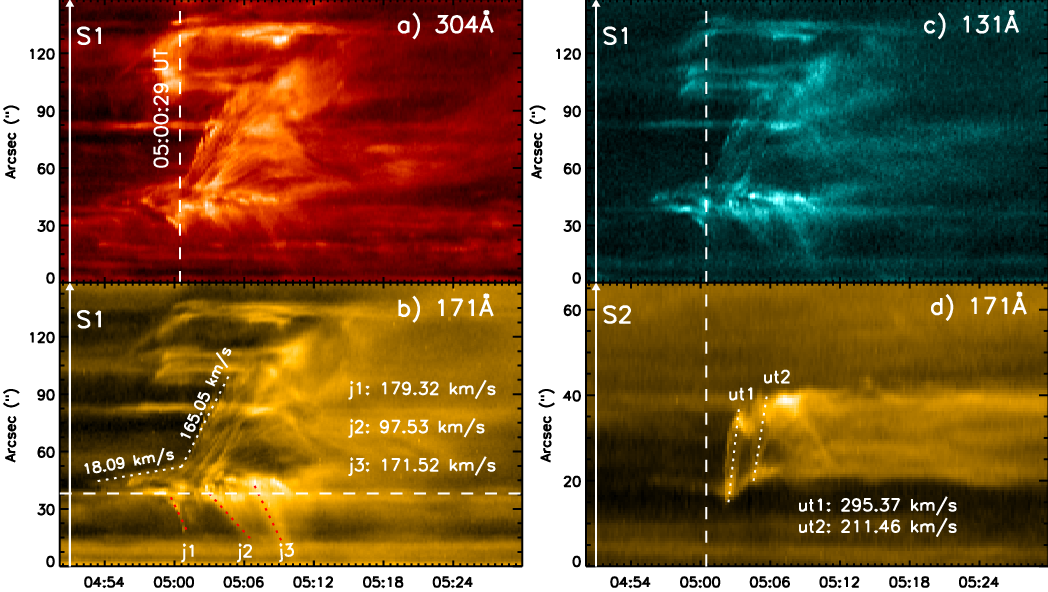}   
\caption{The time-space plots in AIA 304, 171, and 131 \AA\ wavebands. Panels (a)-(c) present the time-slices along slit
 S1, while panel (d) presents the time-slice along slit S2. The positions of these slits are showed in Figure 2c. The
vertical dashed lines indicate the start time of the jet, while the horizontal dashed line denotes the position
of the demarcation point of the jet's spire. The dotted lines in panels (b) and (d) track the mass ejections in
both branches of the jet spire and the transverse motion in the southern branch, respectively, with the measured
projected velocities presented also.
\label{fig3}}
\end{figure}

\begin{figure}
\epsscale{0.95}
\plotone{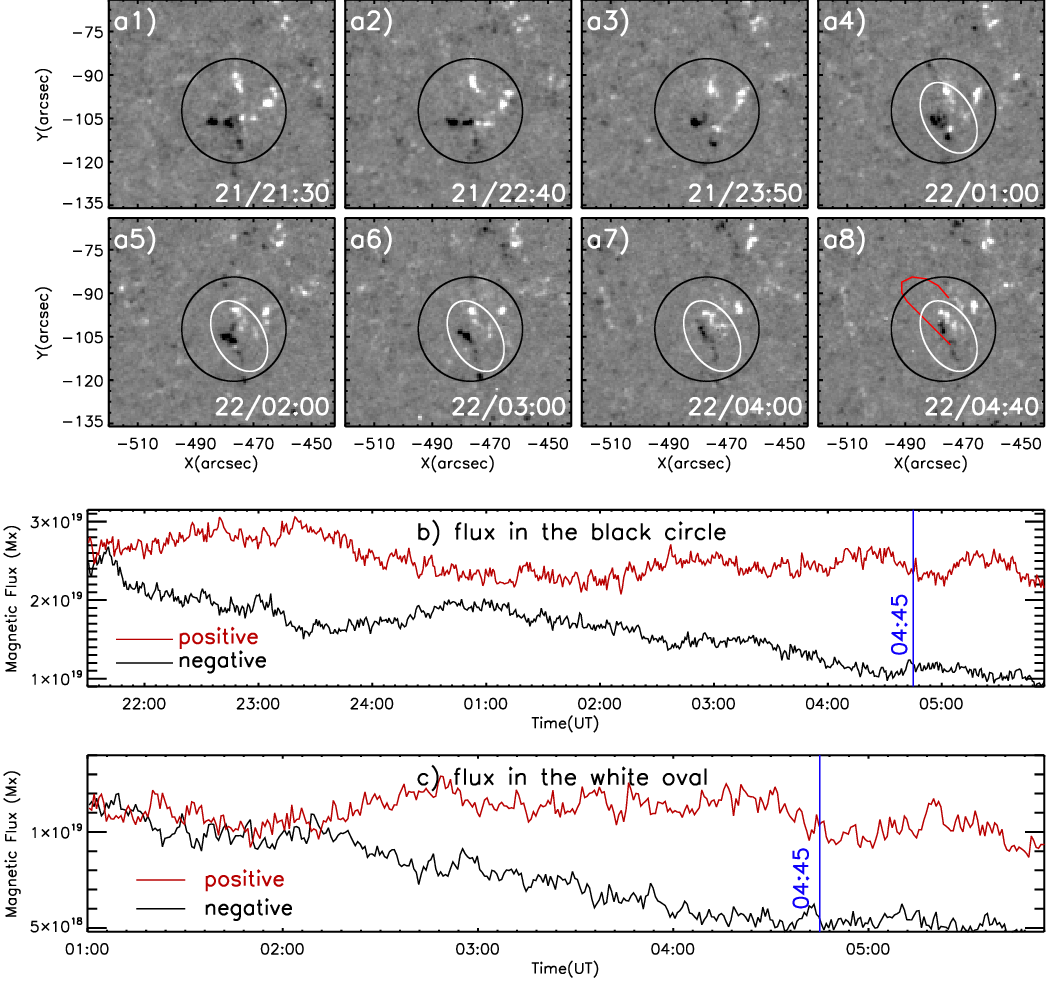}   
\caption{Evolution of the photospheric magnetic field in the source region of the two-sided loop jet.
(a1-a8) HMI LOS magnetograms show the cancellation between the positive and negative magnetic polarities. The red
curve in panel (a8) represents  the outline of F's axis obtained from the simultaneous AIA 304 \AA\ image, and the
FOV is the same as that of the bottom row of Figure 1. (b-c) Unsigned positive and negative magnetic fluxes
calculated in the areas indicated by the black circles and the white ovals in panels (a1-a8). The blue vertical
lines mark the start time of F's eruption. (An animation of HMI LOS magnetograms is available. The animation
lasts 27 seconds, has the same FOV as that of panels (a1)-(a8), and covers the time from 21:30 UT of March 21
to 05:29 UT of March 22. The cadence of the HMI magnetograms is 45s.)
\label{fig4}}
\end{figure}

\begin{figure}
\epsscale{0.95}
\plotone{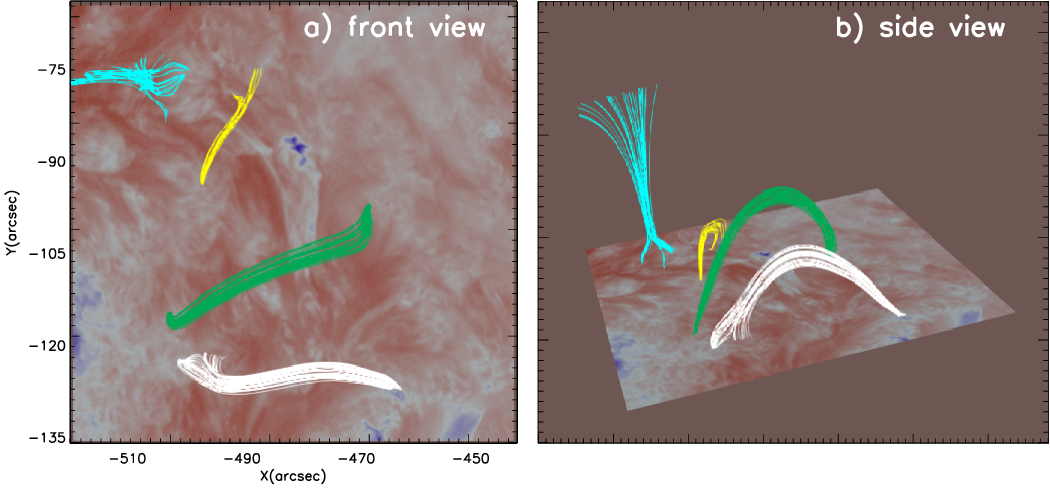}   
\caption{Coronal magnetic field obtained from the local potential field extrapolation explains the asymmetry of the
two-sided loop jet. (a) Front view. (b) Side view. The background image is the 05:09:24 UT AIA 171 \AA\ image, while
the four groups of curves with different colors are the extrapolated 04:00:31 UT coronal magnetic field lines at some
specific positions.
\label{fig5}}
\end{figure}

\end{document}